\documentclass[a4paper,twoside,10pt]{article}

\input{jgrg20.sty}

\usepackage{graphicx}
\usepackage{dcolumn}
\usepackage{amsfonts}
\usepackage{latexsym}
\usepackage{amssymb} 
\usepackage{verbatim}
\usepackage{amsthm}
\usepackage{amsmath}

\newcommand{\RF}{{{\mathbb R}}}

\newtheorem{conjecture}{Conjecture}
\begin{document}
%
\pagestyle{fancy}
\fancyhead{}
  \fancyhead[RO,LE]{\thepage}
  \fancyhead[LO]{K. Nakamura}          
  \fancyhead[RE]{Construction of gauge-invariant variables}   
\rfoot{}
\cfoot{}
\lfoot{}
\label{P47}                             
\title{%
  Construction of gauge-invariant variables for linear-order
  metric perturbations on some background spacetimes
}
%
\author{%
  Kouji Nakamura\footnote{Email address: kouji.nakamura@nao.ac.jp}
}
%
\address{%
  Optical and Infrared Astronomy Division,
  National Astronomical Observatory of Japan,\\
  2-21-1, Osawa, Mitaka, Tokyo 181-8588, Japan
}
%
\abstract{
  Gauge-invariant treatments of general-relativistic
  higher-order perturbations on generic background spacetime is
  proposed.
  We show the fact that the linear-order metric perturbation is
  decomposed into gauge-invariant and gauge-variant parts, which
  was the important premise of this general framework.
  This means that the development the higher-order
  gauge-invariant perturbation theory on generic background
  spacetime is possible.
}

\section{Introduction}


Perturbation theories are powerful techniques in many area of
physics and lead physically fruitful results.
In particular, in general relativity, the construction of exact
solutions is not so easy and known exact solutions are often too
idealized, though there are many known exact solutions to the
Einstein
equation.
Furthermore, in natural phenomena, there always exist
``fluctuations''. 
To describe this, the {\it linear} perturbation theories around
some background spacetime are developed, and are used to
describe fluctuations of our universe, gravity of stars, and
gravitational waves from strongly gravitating sources.


Besides the development of the general-relativistic linear-order
perturbation theory, higher-order general-relativistic
perturbations also have very wide applications, for example,
cosmological
perturbations, black hole perturbations, and perturbation of a
neutron star.
In spite of these applications, there is a delicate issue in
general-relativistic perturbations, which is called 
{\it gauge issue}.
General relativity is based on general covariance. and this
general covariance, the {\it gauge degree of freedom}, which is
an unphysical degree of freedom of perturbations, arises in
general-relativistic perturbations.
To obtain physical results, we have to fix this gauge degree
of freedom or to treat some invariant quantities.
This situation becomes more complicated in higher-order
perturbations.
For this reason, it is worthwhile to investigate higher-order 
gauge-invariant perturbation theory from a general point of
view.


According to this motivation, the general framework of
higher-order general-relativistic gauge-invariant perturbation 
theory has been discussed\cite{kouchan-gauge-inv,kouchan-second}
and applied to cosmological
perturbations\cite{kouchan-cosmo-second,kouchan-second-cosmo-matter}.
This framework is based on a conjecture (Conjecture
\ref{conjecture:decomposition-conjecture} below) which roughly
states that {\it we have already known the procedure to find 
  gauge-invariant variables for a linear-order metric
  perturbations}.
The main purpose of this article is to give the outline of a
proof of this conjecture.


\section{General framework of the higher-order gauge-invariant
  perturbation theory}


In any perturbation theory, we always treat two spacetime
manifolds.
One is the physical spacetime $({\cal M},\bar{g}_{ab})$, which
is our nature itself, and we want to describe 
$({\cal M},\bar{g}_{ab})$ by perturbations. 
The other is the background spacetime $({\cal M}_{0},g_{ab})$,
which is prepared as a reference to calculate perturbations by
us. 
We note that these two spacetimes are distinct.


Further, in any perturbation theory, we write equations for the
perturbation of the variable $Q$ like 
\begin{equation}
  \label{eq:variable-symbolic-perturbation}
  Q(``p\mbox{''}) = Q_{0}(p) + \delta Q(p).
\end{equation}
Equation (\ref{eq:variable-symbolic-perturbation}) gives a
relation between variables on different manifolds.
Actually, $Q(``p\mbox{''})$ in
Eq.~(\ref{eq:variable-symbolic-perturbation}) is a variable on 
${\cal M}$, while $Q_{0}(p)$ and $\delta Q(p)$ are variables on
${\cal M}_{0}$.
Further, since regard
Eq.~(\ref{eq:variable-symbolic-perturbation}) as a field
equation, this is an implicit assumption of the existence of a
point identification map 
${\cal M}_{0}\rightarrow{\cal M}$ $:$ $p\in{\cal M}_{0}\mapsto
``p\mbox{''}\in{\cal M}$.
This identification map is a {\it gauge choice} in
perturbation theories\cite{J.M.Stewart-M.Walker11974}.


To develop this understanding of the ``gauge'', we introduce an
infinitesimal parameter $\lambda$ and $(n+1)+1$-dimensional
manifold ${\cal N}={\cal M}\times\RF$ ($n+1=\dim{\cal M}$) so
that ${\cal M}_{0}=\left.{\cal N}\right|_{\lambda=0}$ and 
${\cal M}={\cal M}_{\lambda}=\left.{\cal N}\right|_{\RF=\lambda}$.
On ${\cal N}$, the gauge choice is regarded as a diffeomorphism  
${\cal X}_{\lambda}:{\cal N}\rightarrow{\cal N}$ such that
${\cal X}_{\lambda}:{\cal M}_{0}\rightarrow{\cal M}_{\lambda}$.
Further, we introduce a gauge choice ${\cal X}_{\lambda}$ as an
exponential map with a generator ${}^{{\cal X}}\!\eta^{a}$
which is chosen so that its integral curve in ${\cal N}$ is
transverse to each ${\cal M}_{\lambda}$ everywhere on 
${\cal N}$.
Points lying on the same integral curve are regarded as the
``same'' by the gauge choice ${\cal X}_{\lambda}$.


The first- and the second-order perturbations of
the variable $Q$ on ${\cal M}_{\lambda}$ are defined by the
pulled-back ${\cal X}_{\lambda}^{*}Q$ on ${\cal M}_{0}$ induced
by ${\cal X}_{\lambda}$, and expanded as 
\begin{eqnarray}
  {\cal X}_{\lambda}^{*}Q
  =
  Q_{0}
  + \lambda \left.{\pounds}_{{}^{{\cal X}}\!\eta}Q\right|_{{\cal M}_{0}}
  + \frac{1}{2} \lambda^{2} 
  \left.{\pounds}_{{}^{{\cal X}}\!\eta}^{2}Q\right|_{{\cal M}_{0}}
  + O(\lambda^{3}),
  \label{eq:perturbative-expansion-of-Q-def}
\end{eqnarray}
$Q_{0}=\left.Q\right|_{{\cal M}_{0}}$ is the background value of
$Q$ and all terms in
Eq.~(\ref{eq:perturbative-expansion-of-Q-def}) are evaluated on
${\cal M}_{0}$.
Since Eq.~(\ref{eq:perturbative-expansion-of-Q-def}) is just the
perturbative expansion of ${\cal X}^{*}_{\lambda}Q_{\lambda}$,
the first- and the second-order perturbations of $Q$ are given
by  
${}^{(1)}_{{\cal X}}\!Q:=\left.{\pounds}_{{}^{{\cal X}}\!\eta}Q\right|_{{\cal M}_{0}}$
and 
${}^{(2)}_{{\cal X}}\!Q:=\left.{\pounds}_{{}^{{\cal X}}\!\eta}^{2}Q\right|_{{\cal M}_{0}}$,
respectively.


When we have two gauge choices ${\cal X}_{\lambda}$ and 
${\cal Y}_{\lambda}$ with the generators ${}^{{\cal X}}\!\eta^{a}$
and ${}^{{\cal Y}}\!\eta^{a}$, respectively, and when these
generators have the different tangential components 
to each ${\cal M}_{\lambda}$, ${\cal X}_{\lambda}$ and 
${\cal Y}_{\lambda}$ are regarded as {\it different gauge choices}.
The {\it gauge-transformation} is regarded as the change of the
gauge choice ${\cal X}_{\lambda}\rightarrow{\cal Y}_{\lambda}$,
which is given by the diffeomorphism 
$\Phi_{\lambda}:=\left({\cal X}_{\lambda}\right)^{-1}\circ{\cal Y}_{\lambda}
  : {\cal M}_{0} \rightarrow {\cal M}_{0}$.
The diffeomorphism $\Phi_{\lambda}$ does change the point
identification.
$\Phi_{\lambda}$ induces a pull-back from the representation
${\cal X}_{\lambda}^{*}\!Q_{\lambda}$ to the representation
${\cal Y}_{\lambda}^{*}\!Q_{\lambda}$ as 
${\cal Y}_{\lambda}^{*}\!Q_{\lambda}=\Phi_{\lambda}^{*}{\cal X}_{\lambda}^{*}\!Q_{\lambda}$.
From general arguments of the Taylor
expansion, the pull-back $\Phi_{\lambda}^{*}$ is expanded as
\begin{eqnarray}
  {\cal Y}_{\lambda}^{*}\!Q_{\lambda}
  =
  {\cal X}_{\lambda}^{*}\!Q_{\lambda}
  + \lambda {\pounds}_{\xi_{(1)}} {\cal X}_{\lambda}^{*}\!Q_{\lambda}
  + \frac{1}{2} \lambda \left(
    {\pounds}_{\xi_{(2)}} + {\pounds}_{\xi_{(1)}}^{2}
  \right) {\cal X}_{\lambda}^{*}\!Q_{\lambda}
  + O(\lambda^{3}),
  \label{eq:Bruni-46-one}
\end{eqnarray}
where $\xi_{(1)}^{a}$ and $\xi_{(2)}^{a}$ are the generators of
$\Phi_{\lambda}$.
From Eqs.~(\ref{eq:perturbative-expansion-of-Q-def}) and
(\ref{eq:Bruni-46-one}), each order gauge-transformation is
given as
\begin{eqnarray}
  \label{eq:Bruni-47-one}
  {}^{(1)}_{\;{\cal Y}}\!Q - {}^{(1)}_{\;{\cal X}}\!Q = 
  {\pounds}_{\xi_{(1)}}Q_{0}
  ,
  \quad
  {}^{(2)}_{\;\cal Y}\!Q - {}^{(2)}_{\;\cal X}\!Q = 
  2 {\pounds}_{\xi_{(1)}} {}^{(1)}_{\;\cal X}\!Q 
  +\left\{{\pounds}_{\xi_{(2)}}+{\pounds}_{\xi_{(1)}}^{2}\right\} Q_{0}.
\end{eqnarray}
We also employ the {\it order by order gauge invariance} as a
concept of gauge invariance\cite{kouchan-second-cosmo-matter}. 
We call the $k$th-order perturbation ${}^{(p)}_{{\cal X}}\!Q$ is
gauge invariant iff ${}^{(k)}_{\;\cal X}\!Q = {}^{(k)}_{\;\cal Y}\!Q$
for any gauge choice ${\cal X}_{\lambda}$ and
${\cal Y}_{\lambda}$.


Based on the above set up, we proposed a procedure to construct
gauge-invariant variables of higher-order
perturbations\cite{kouchan-gauge-inv}.
First, we expand the metric on the physical spacetime 
${\cal M}_{\lambda}$, which is pulled back to the background
spacetime ${\cal M}_{0}$ through a gauge choice 
${\cal X}_{\lambda}$ as ${\cal X}^{*}_{\lambda}\bar{g}_{ab}$ $=$
$g_{ab}$ $+$ $\lambda {}_{{\cal X}}\!h_{ab}$ $+$
$\frac{\lambda^{2}}{2}{}_{{\cal X}}\!l_{ab}$ $+$ $O^{3}(\lambda)$.
Although this expression of metric perturbations depends
entirely on the gauge choice ${\cal X}_{\lambda}$, henceforth,
we do not explicitly express the index of the gauge choice
${\cal X}_{\lambda}$ in the expression if there is no
possibility of confusion. 
The important premise of our proposal was the following
conjecture\cite{kouchan-gauge-inv} for $h_{ab}$ : 
\begin{conjecture}
  \label{conjecture:decomposition-conjecture}
  For a second-rank tensor $h_{ab}$, whose gauge transformation
  is given by (\ref{eq:Bruni-47-one}), there exist a tensor
  ${\cal H}_{ab}$ and a vector $X^{a}$ such that $h_{ab}$ is
  decomposed as
  \begin{eqnarray}
    h_{ab} =: {\cal H}_{ab} + {\pounds}_{X}g_{ab},
    \label{eq:linear-metric-decomp}
  \end{eqnarray}
  where ${\cal H}_{ab}$ and $X^{a}$ are transformed as
  \begin{equation}
    {}_{{\cal Y}}\!{\cal H}_{ab} - {}_{{\cal X}}\!{\cal H}_{ab} =  0, 
    \quad
    {}_{\quad{\cal Y}}\!X^{a} - {}_{{\cal X}}\!X^{a} = \xi^{a}_{(1)} 
    \label{eq:linear-metric-decomp-gauge-trans}
  \end{equation}
  under the gauge transformation (\ref{eq:Bruni-47-one}),
  respectively.
\end{conjecture}
We call ${\cal H}_{ab}$ and $X^{a}$ are the 
{\it gauge-invariant part} and the {\it gauge-variant part} 
of $h_{ab}$, respectively.


Although Conjecture \ref{conjecture:decomposition-conjecture} is
nontrivial on generic background spacetime, once we accept this
conjecture, we can always find gauge-invariant variables for
higher-order perturbations\cite{kouchan-gauge-inv}. 
Using Conjecture \ref{conjecture:decomposition-conjecture}, the
second-order metric perturbation $l_{ab}$ is decomposed as
\begin{eqnarray}
  \label{eq:H-ab-in-gauge-X-def-second-1}
  l_{ab}
  =:
  {\cal L}_{ab} + 2 {\pounds}_{X} h_{ab}
  + \left(
      {\pounds}_{Y}
    - {\pounds}_{X}^{2} 
  \right)
  g_{ab},
\end{eqnarray}
where ${}_{{\cal Y}}\!{\cal L}_{ab}-{}_{{\cal X}}\!{\cal L}_{ab}=0$  
and 
${}_{{\cal Y}}\!Y^{a}-{}_{{\cal X}}\!Y^{a}=\xi_{(2)}^{a}+[\xi_{(1)},X]^{a}$.
Furthermore, using the first- and second-order gauge-variant
parts, $X^{a}$ and $Y^{a}$, of the metric perturbations,
gauge-invariant variables for an arbitrary tensor field $Q$ 
other than the metric can be defined by
\begin{eqnarray}
  \label{eq:matter-gauge-inv-def-1.0}
  {}^{(1)}\!{\cal Q} &:=& {}^{(1)}\!Q - {\pounds}_{X}Q_{0}
  , \quad
  {}^{(2)}\!{\cal Q} := {}^{(2)}\!Q - 2 {\pounds}_{X} {}^{(1)}Q 
  - \left\{ {\pounds}_{Y} - {\pounds}_{X}^{2} \right\} Q_{0}
  .
\end{eqnarray}
These definitions (\ref{eq:matter-gauge-inv-def-1.0}) also imply
that any perturbation of first and second order is always
decomposed into gauge-invariant and gauge-variant parts.
These decomposition formulae are
universal\cite{kouchan-second-cosmo-matter,kouchan-second}.
Further, when we impose order by order equations for the
perturbations, any perturbative equations are automatically
given in gauge-invariant
form\cite{kouchan-second-cosmo-matter,kouchan-second}.


Thus, based only on Conjecture
\ref{conjecture:decomposition-conjecture}, we have developed
the general framework of second-order general relativistic
perturbation theory without detail information of the background
metric $g_{ab}$.


\section{Decomposition of the linear-order metric perturbation}


Now, we show the outline of a proof of Conjecture
\ref{conjecture:decomposition-conjecture}.
To do this, we only consider the background spacetimes which
admit ADM decomposition.
Therefore, the background spacetime ${\cal M}_{0}$ considered
here is $n+1$-dimensional spacetime which is described by the
direct product $\RF\times\Sigma$.
Here, $\RF$ is a time direction and $\Sigma$ is the spacelike
hypersurface ($\dim\Sigma = n$).
The background metric $g_{ab}$ is given as
\begin{eqnarray}
  \label{eq:gdb-decomp-dd-minus-main}
  g_{ab} \! = \! - \alpha^{2} (dt)_{a} (dt)_{b}
  + q_{ij}
  (dx^{i} + \beta^{i}dt)_{a}
  (dx^{j} + \beta^{j}dt)_{b}.
\end{eqnarray}
In this article, we only consider the case where
$\alpha = 1$ and $\beta^{i} = 0$, for simplicity.
The proof shown here is extended to general
case\cite{kouchan-in-preparation}.


To consider the decomposition (\ref{eq:linear-metric-decomp}) of
$h_{ab}$, first, we consider the components of the metric
$h_{ab}$ as $h_{ab}$ $=:$ $h_{tt}(dt)_{a}(dt)_{b}$ $+$
$2h_{ti}(dt)_{(a}(dx^{i})_{b)}$ $+$ $h_{ij}(dx^{i})_{a}(dx^{j})_{b}$.
Under the gauge-transformation (\ref{eq:Bruni-47-one}), these
components $\{h_{tt},h_{ti},h_{ij}\}$ are transformed as
\begin{eqnarray}
  \label{eq:gauge-trans-of-htt-ADM-BG-case2}
  {}_{{\cal Y}}h_{tt}
  -
  {}_{{\cal X}}h_{tt}
  =
  2 \partial_{t}\xi_{t}
  , \quad
  {}_{{\cal Y}}h_{ti}
  -
  {}_{{\cal X}}h_{ti}
  =
  \partial_{t}\xi_{i}
  + D_{i}\xi_{t}
  + 2 K^{j}_{\;\;i} \xi_{j}
  , \quad
  {}_{{\cal Y}}h_{ij}
  -
  {}_{{\cal X}}h_{ij}
  =
  2 D_{(i}\xi_{j)}
  + 2 K_{ij} \xi_{t}
  .
\end{eqnarray}
where $K_{ij}$ is the extrinsic curvature of $\Sigma$ and
$D_{i}$ is the covariant derivative associate with the metric
$q_{ij}$ ($D_{i}q_{jk}=0$). 
In our case, $K_{ij}=-\frac{1}{2} \partial_{t}q_{ij}$.
Inspecting gauge-transformation rules
(\ref{eq:gauge-trans-of-htt-ADM-BG-case2}), we introduce a new
symmetric tensor $\hat{H}_{ab}$ whose components are given by 
$\hat{H}_{tt}$ $:=$ $h_{tt}$, $\hat{H}_{ti}$ $:=$ $h_{ti}$,
$\hat{H}_{ij}$ $:=$ $h_{ij}$ $-$ $2K_{ij}X_{t}$.
Here, we assume the existence of the variable $X_{t}$
whose gauge-transformation rule is given by 
${}_{{\cal Y}}X_{t}-{}_{{\cal X}}X_{t}=\xi_{t}$.
This assumption is confirmed later soon. 
Since the components $\hat{H}_{ti}$ and $\hat{H}_{ij}$ are
a vector and a symmetric tensor on $\Sigma$, respectively,
$\hat{H}_{ti}$ and $\hat{H}_{ij}$ are decomposed as\cite{J.W.York-1973}
\begin{eqnarray}
  \label{eq:K.Nakamura-2010-2-simple-4-7}
  \hat{H}_{ti} &=& D_{i}h_{(VL)} + h_{(V)i}, \;\; D^{i}h_{(V)i} = 0,
  \\
  \label{eq:K.Nakamura-2010-2-simple-4-8}
  \hat{H}_{ij} &=& \frac{1}{n} q_{ij} h_{(L)}
  + 2 \left(D_{(i}h_{(TV)j)} - \frac{1}{n}q_{ij}D^{l}h_{(TV)l}\right)
  + h_{(TT)ij}, \;\; D^{i}h_{(TT)ij} = 0,
  \\
  \label{eq:K.Nakamura-2010-2-simple-4-10}
  h_{(TV)i} &=& D_{i}h_{(TVL)} + h_{(TVV)i}, \;\;
  D^{i}h_{(TVV)i} = 0.
\end{eqnarray}
The one-to-one correspondence between
$\{\hat{H}_{ti}$, $\hat{H}_{ij}\}$ and
$\{h_{(VL)}$,$h_{(V)i}$,$h_{(L)}$,$h_{(TVL)}$,$h_{(TVV)i}$,$h_{(TT)ij}\}$
is guaranteed by the existence of the Green functions of operators
$\Delta:=D^{i}D_{i}$ and 
${\cal D}^{ij}:=q^{ij}\Delta+\left(1-\frac{2}{n}\right)D^{i}D^{j}+{}^{(n)}\!R^{ij}$,
where ${}^{(n)}\!R^{ij}$ is the Ricci curvature on $\Sigma$.
Here, we assume their existence.
Gauge-transformation rules for $\{h_{tt}$, $h_{(VL)}$,
$h_{(V)i}$, $h_{(L)}$, $h_{(TVL)}$, $h_{(TVV)i}$, $h_{(TT)ij}\}$ 
are summarized as
\begin{eqnarray}
  \!\!\!
  {}_{{\cal Y}}h_{tt}
  -
  {}_{{\cal X}}h_{tt}
  \!\!\!&=&\!\!\!
  2 \partial_{t}\xi_{t}
  \label{eq:K.Nakamura-2010-2-simple-4-39-2}
  , \quad
  {}_{{\cal Y}}h_{(TT)ij} - {}_{{\cal X}}h_{(TT)ij}
  =
  0
  , \\
  \!\!\!
  {}_{{\cal Y}}h_{(VL)} - {}_{{\cal X}}h_{(VL)}
  \!\!\!&=&\!\!\!
    \partial_{t}\xi_{(L)}
  + \xi_{t}
  +
  \Delta^{-1}
  \left[
    2 D_{i}\left( K^{ij} D_{j}\xi_{(L)} \right)
    + D^{k}K \xi_{(V)k}
  \right]
  \label{eq:K.Nakamura-2010-2-simple-4-40}
  , \\
  \!\!\!
  {}_{{\cal Y}}h_{(V)i}
  -
  {}_{{\cal X}}h_{(V)i}
  \!\!\!&=&\!\!\!
    \partial_{t}\xi_{(V)i}
  + 2 K^{j}_{\;\;i} D_{j}\xi_{(L)}
  + 2 K^{j}_{\;\;i} \xi_{(V)j}
  - D_{i}\Delta^{-1}
  \left[
    2 D_{i} \left( K^{ij} D_{j}\xi_{(L)} \right)
    + D^{k}K \xi_{(V)k}
  \right]
  \label{eq:K.Nakamura-2010-2-simple-4-41}
  ,
  \\
  \!\!\!
  {}_{{\cal Y}}h_{(L)}
  -
  {}_{{\cal X}}h_{(L)}
  \!\!\!&=&\!\!\!
  2 D^{i}\xi_{i}
  , \quad
  {}_{{\cal Y}}h_{(TVL)} - {}_{{\cal X}}h_{(TVL)}
  =
  \xi_{(L)}
  , \quad
  {}_{{\cal Y}}h_{(TVV)l} - {}_{{\cal X}}h_{(TVV)l}
  =
  \xi_{(V)l}
  , \quad
  .
  \label{eq:K.Nakamura-2010-2-simple-4-44}
\end{eqnarray}
where we decompose $\xi_{i}=:D_{i}\xi_{(L)}+\xi_{(V)i}$,
$D^{i}\xi_{(V)i}=0$. 


We first find the variable $X_{t}$ in the definition of
$\hat{H}_{ab}$. 
From the above gauge-transformation rules, we see that
the combination $X_{t}$ $:=$ $h_{(VL)}$ $-$
$\partial_{t}h_{(TVL)}$ $-$
$\Delta^{-1}\left[2D_{i}\left(K^{ij}D_{j}h_{(TVL)}\right)+D^{k}K h_{(TVV)k} \right]$
satisfy ${}_{{\cal Y}}X_{t}-{}_{{\cal X}}X_{t}=\xi_{t}$. 
We also find the variable $X_{i}$ $:=$ $h_{(TV)i}$ $=$
$D_{i}h_{(TVL)}$ $+$ $h_{(TVV)i}$ satisfy the
gauge-transformation rule ${}_{{\cal Y}}X_{i}-{}_{{\cal X}}X_{i}=\xi_{i}$.


Inspecting gauge-transformation rules
(\ref{eq:K.Nakamura-2010-2-simple-4-39-2})--(\ref{eq:K.Nakamura-2010-2-simple-4-44})
and using the variables $X_{t}$ and $X_{i}$,
we find gauge-invariant variables as follows:
\begin{eqnarray}
  \label{eq:K.Nakamura-2010-2-simple-4-58}
  - 2 \Phi &:=& h_{tt} - 2 \partial_{t}\hat{X}_{t}, \quad
  - 2 n \Psi := h_{(L)} - 2 D^{i}\hat{X}_{i}, \quad
  \chi_{ij} := h_{(TT)ij}
  , \\
  \nu_{i}
  &:=&
  h_{(V)i}
  - \partial_{t}h_{(TVV)i}
  - 2 K^{j}_{\;\;i} \left( D_{j}h_{(TVL)} + h_{(TVV)j} \right)
  \nonumber\\
  &&
  + D_{i}\Delta^{-1}
  \left[
    2 D_{i}\left( K^{ij} D_{j}h_{(TVL)} \right)
    + D^{k}K h_{(TVV)k}
  \right]
  .
\end{eqnarray}
Actually, it is straightforward to confirm the gauge-invariance
of these variables.
In terms of the variables $\Phi$, $\Psi$, $\nu_{i}$,
$\chi_{ij}$, $X_{t}$, and $X_{i}$, original components of
$h_{ab}$ is given by 
\begin{eqnarray}
  \label{eq:K.Nakamura-2010-2-simple-4-71}
  h_{tt} &=& - 2 \Phi + 2 \partial_{t}X_{t}, \quad
  h_{ti}
  =
    \nu_{i}
  + D_{i}X_{t}
  + \partial_{t}X_{i}
  + 2 K^{j}_{\;\;i} X_{j}
  , \\
  h_{ij}
  &=&
  - 2 \Psi q_{ij} 
  + \chi_{ij}
  + D_{i}X_{j} + D_{j}X_{i}
  + 2 K_{ij} X_{t}
  \label{eq:K.Nakamura-2010-2-simple-4-73}
  .
\end{eqnarray}
Comparing Eq.~(\ref{eq:linear-metric-decomp}),
a natural choice of ${\cal H}_{ab}$ and $X_{a}$ are
\begin{eqnarray}
  \label{eq:calHab-component-identification-case2}
  {\cal H}_{ab}
  = - 2 \Phi (dt)_{a}(dt)_{b}
  + 2 \nu_{i} (dt)_{(a}(dx^{i})_{b)}
  + \left(- 2 \Psi q_{ij} + \chi_{ij}\right) (dx^{i})_{a} (dx^{i})_{b}
  , \quad
  X_{a} = X_{t}(dt)_{a} + X_{i} (dx^{i}).
\end{eqnarray}
These show that the linear-order metric perturbation
$h_{ab}$ is decomposed into the form
Eq.~(\ref{eq:linear-metric-decomp}).


\section{Discussion}


In our proof, we assumed the existence of the Green functions for
the derivative operators $\Delta$ and ${\cal D}^{ij}$.
This implies that we have ignored the modes which belong to the
kernel of these derivative operators.
To includes these modes into our consideration, different
treatments of perturbations will be necessary.
We call this problem as {\it zero-mode problem}.
We leave this zero-mode problem as a future work.


Although this zero-mode problem should be resolved, we confirmed 
the important premise of our general framework of second-order
gauge-invariant perturbation theory on generic background
spacetime.
This means that we have the possibility of applications of our
framework for the second-order gauge-invariant perturbation
theory to perturbations on generic background spacetime.
Furthermore, the similar development will be also possible for
the any order perturbation in two-parameter
case\cite{kouchan-gauge-inv}.
Thus, we may say that wide applications of our gauge-invariant
perturbation theory are opened.
We also leave these developments as future works.


The author deeply acknowledged to Professor Masa-Katsu Fujimoto
in National Astronomical Observatory of Japan for his various
support.



\end{document}